\documentclass[aps,pra,twocolumn,showpacs,groupedaddress]{revtex4-1}
\usepackage{graphicx}
\usepackage{epstopdf} 
\begin{document}

\title{Theory of lasing in a two-dimensional array of plasmonic nanolasers}

\author{V.G.~Bordo}
\email{bordo@sdu.dk}

\affiliation{SDU Electrical Engineering, University of Southern Denmark,
Alsion 2, DK-6400 S{\o}nderborg, Denmark}


\date{\today}

\begin{abstract}
A theory of lasing in a two-dimensional array of metal nanoparticles (MNPs) covered with a thin layer of fluorescent molecules is developed from first principles. The approach is based on a rigorous account of the local field in a close vicinity of a reflective surface which provides a feedback for molecular dipole oscillations. The theory predicts the lasing threshold in such an open cavity in terms of the polar angle of laser emission, MNPs shape and the molecular layer thickness. It is demonstrated that the latter parameter plays a crucial role in the lasing condition and controls a switching from conventional lasing to lasing without inversion. This research is inspired by recent experiments in this field [N. Toropov et al, Adv. Photonics Res. {\bf 2}, 2000083 (2021)] and provides the numerical calculations carried out for the experimental conditions.
\end{abstract}


\maketitle

\section{Introduction}
The current trend in nanophotonics requires miniaturization of coherent light sources \cite{Odom17,Zhang20}. Nanoscale lasers (nanolasers) are promising for diverse applications, from integration in optical data networks for increasing optical
communication speeds to opening new possibilities in bioimaging and ultra-sensitive chemical analysis.\\
While conventional photonic lasers suffer from the diffraction limit which prevents their reduction in size, the so-called plasmonic nanolasers and spasers \cite{Odom17,Zhang20,Stockman03,Hill07,Noginov09,Oulton09}, which exploit strongly localized electromagnetic field of surface plasmon polaritons, allow one to reach the ultimately small dimensions.\\
The active molecules which constitute gain medium in nanolasers are confined within a volume with subwavelength dimensions that highlights the quantum electrodynamical effects near an interface or in a cavity \cite{Berman94,Walter06} which are hidden in macroscopic lasers. A molecule emitting light in a close vicinity of a surface undergoes the backaction from it due to the reflected field. The phase of this backaction depends on the molecule-surface distance that leads, in particular, to the oscillating variation of the molecule relaxation time with distance \cite{Drexhage68,Chance74}.\\
In a large ensemble of active molecules the backaction scales with their number that can significantly modify the molecular polarization behavior. In such a case the backaction phase is averaged over the ensemble and depends on the gain medium location and size. A careful account of this effect is crucial for a proper analysis of the nanolaser and spaser dynamics. It is based on the dyadic Green's function approach and was performed before for nanowire lasers \cite{Bordo13} and core-shell nanoparticle spasers \cite{Pustovit16,Cuerda16,Shahbazyan17,Bordo17a}. \\
In the present paper, we report on a dramatic impact which the backaction has on the laser operation in a two-dimensional array of metal nanoparticles (MNPs) in contact with molecular gain medium. We theoretically investigate lasing in a model system which resembles the configuration realized in recent experiments \cite{Toropov21} where $p$-polarized laser emission was observed. We demonstrate that the lasing condition is determined not just by the population inversion in gain medium as in conventional lasers, but also by the averaged phase factor which stems from the backaction of the MNPs layer. As a result, the structure can lase either with or without population inversion, depending on the molecular layer thickness.\\ 
The phenomenon of lasing without inversion is based on a phase-sensitive backaction from MNPs which support localized surface plasmons (LSPs). It was predicted before for quite diverse systems: semiconductor nanowire lasers \cite{Bordo13}, surface plasmon polaritons between two metal surfaces \cite{Bordo16} and Rydberg atoms in a beam propagating near a metal surface \cite{Bordo17}. The same mechanism governs the generation of LSPs in the structure (spasing), however the analysis of the spaser dynamics is beyond the scope of the present paper. In this sense the system under discussion falls into a large class of structures called "lasing spasers" \cite{Zheludev08}.\\
The paper is organized as follows. Section \ref{sec:model} introduces the theoretical model which is used to describe the system and calculate the molecular polarization in gain medium. In Sec. \ref{sec:lasing} we derive the lasing condition which is numerically analyzed in Sec. \ref{sec:numerical}. The main results of the paper are summarized in Sec. \ref{sec:conclusion}.
\section{Theoretical model}\label{sec:model}
\subsection{System under consideration}
\begin{figure}
\includegraphics[width=\linewidth]{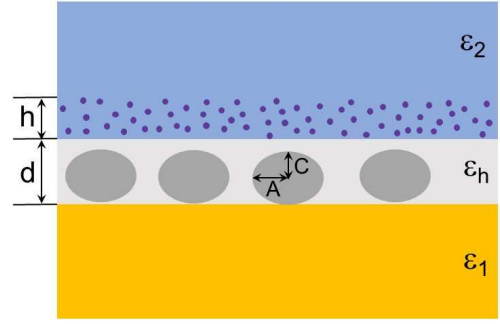}
\caption{\label{fig:sketch} The sketch of the structure under consideration.}
\end{figure}

Let us consider an array of MNPs randomly distributed within a dielectric host material of thickness $d$ and having the dielectric function $\epsilon_h$ which is disposed on a dielectric substrate with the dielectric function $\epsilon_1$. We direct the $z$ coordinate axis along the normal to the slab boundaries and assume that MNPs have a shape of spheroids with their axes of rotational symmetry (semi-axis $C$) being parallel to the $z$ axis. The optical response of MNPs is described by the Drude dielectric function
\begin{equation}\label{eq:drude}
\epsilon_m(\omega)=\epsilon_{\infty}-\frac{\omega_p^2}{\omega(\omega+i\Gamma)},
\end{equation}
where $\omega=2\pi c/\lambda$ is the frequency of the incident light with $c$ being the speed of light in vacuum and $\lambda$ being the wavelength, $\epsilon_{\infty}$ is the offset originating from the interband transitions, $\omega_p$ is the metal plasma frequency and $\Gamma$ is the relaxation constant.\\
Assuming that both the MNPs size and the slab thickness are much smaller that the wavelength ($C,d\ll\lambda$) one can use the quasistatic approximation and find the dielectric tensor components of the slab containing MNPs as follows ($i,j=x,y,z$)
\begin{eqnarray}\label{eq:slab}
\epsilon_{ii}(\omega)=\epsilon_h+4\pi N\alpha_{ii}(\omega),\\
\epsilon_{ij}(\omega)=0,\quad i\neq j,
\end{eqnarray}
where
\begin{equation}\label{eq:alpha}
\alpha_{ii}(\omega)=\frac{1}{3}A^2C\epsilon_h\frac{\epsilon_m(\omega)-\epsilon_h}{\epsilon_h+L_i(\xi)[\epsilon_m(\omega)-\epsilon_h]}
\end{equation}
is the spheroid polarizability components with $A$ being the semi-axis perpendicular to the $z$ axis and $\xi=A/C$ being the aspect ratio \cite{Landau}, $N$ is the volume number density of MNPs. The depolarization coefficients $L_i(\xi)$ are defined as follows
\begin{equation}
L_z(\xi)=\frac{1+e^2(\xi)}{e^3(\xi)}[e(\xi)-\arctan e(\xi)]
\end{equation}
and
\begin{equation}\label{eq:Lx}
L_x(\xi)=L_y(\xi)=\frac{1}{2}[1-L_z(\xi)]
\end{equation}
with $e(\xi)=\sqrt{\xi^2-1}$. When writing Eq. (\ref{eq:slab}) we neglected the effect of the local field in the slab which is estimated to be of the order of $d/\lambda$. Let us note that this equation can be rewritten in terms of the volume fraction of MNPs, $f=(4\pi/3)A^2CN$, as follows
\begin{equation}\label{eq:slab_f}
\epsilon_{ii}(\omega)=\epsilon_h\left\{1+f\frac{\epsilon_m(\omega)-\epsilon_h}{\epsilon_h+L_i(\xi)[\epsilon_m(\omega)-\epsilon_h]}\right\}.
\end{equation}
Equation (\ref{eq:Lx}) entails that $\alpha_{xx}(\omega)=\alpha_{yy}(\omega)$ and hence the slab is optically uniaxial with the optical axis being parallel to the $z$ axis. The ordinary and extraordinary refractive indices are determined by the dielectric tensor components $\epsilon_o(\omega)\equiv \epsilon_{xx}(\omega)=\epsilon_{yy}(\omega)$ and $\epsilon_e(\omega)\equiv \epsilon_{zz}(\omega)$, respectively.\\
The minima of the denominators in Eq. (\ref{eq:alpha}) which occur when
\begin{equation}
\text{Re}[\epsilon_m(\omega)]=-\epsilon_h\tau_i(\xi)
\end{equation}
with $\tau_i(\xi)=[1-L_i(\xi)]/L_i(\xi)$ signify excitation of LSPs in MNPs when their optical response is maximal. The LSP frequencies can be deduced from Eqs. (\ref{eq:drude}) and (\ref{eq:alpha}) and are found as follows
\begin{equation}
\omega_i=\frac{\omega_p}{\sqrt{\epsilon_{\infty}+\epsilon_h\tau_i(\xi)}},
\end{equation}
the frequencies $\omega_x$ and $\omega_y$ being equal to each other. For a spherical MNP $(\xi=1)$ $L_x=L_y=L_z=1/3$ and all three frequencies coincide with each other.\\
Let us assume further that a semi-infinite dielectric medium disposed above the slab with MNPs has the dielectric function $\epsilon_2$ and contains a layer of active molecules of thickness $h$ adjacent to the slab. We assume also that one of the molecular transitions has the frequency $\omega_0$ which is close to one of the LSP frequencies and that this transition can be optically pumped by an external source of radiation. Figure \ref{fig:sketch} shows the sketch of the system under discussion. \\
\subsection{Evolution of the molecular polarization}
The radiation emitted from the system can be regarded as being originating from the molecular dipoles ${\bf p}({\bf r},t)$, ${\bf r}$ being the radius vector of the molecule, oscillating near the molecular transition frequency. The electromagnetic field which exists in the system can be decomposed into the negative and positive frequency parts as
\begin{equation}
{\bf E}(t)={\bf E}^{(-)}(t)e^{i\omega t}+{\bf E}^{(+)}(t)e^{-i\omega t}
\end{equation}
with
\begin{equation}
{\bf E}^{(\pm)}(t)=[{\bf E}^{(\mp)}(t)]^*
\end{equation}
being the slowly varying amplitudes in which we have omitted the dependence on the molecule coordinates for the sake of brevity.\\
The molecular polarization ${\bf P}({\bf r},t)=\mathcal{N}{\bf p}({\bf r},t)$ with $\mathcal{N}$ being the number density of active molecules takes a similar form,
\begin{equation}
{\bf P}(t)={\bf P}^{(-)}(t)e^{i\omega t}+{\bf P}^{(+)}(t)e^{-i\omega t},
\end{equation}
where the slow amplitudes ${\bf P}^{(\pm)}(t)$ satisfy the optical Bloch equations \cite{Haken}
\begin{eqnarray}\label{eq:bloch1}
\frac{\partial {\bf P}^{(+)}(t)}{\partial t}=-(\gamma_{\perp }-i\Delta){\bf P}^{(+)}(t)\nonumber\\
-\frac{i}{3\hbar } \mu^2D(t){\bf E}^{(+)}(t),
\end{eqnarray}
\begin{eqnarray}\label{eq:bloch2}
\frac{\partial D(t)}{\partial t}=-\gamma_{\parallel }\left[D(t)-D_0\right]\nonumber\\
+\frac{2i}{\hbar }\left[{\bf P}^{(-)}(t){\bf E}^{(+)}(t)-{\bf P}^{(+)}(t){\bf E}^{(-)}(t)\right].
\end{eqnarray}
Here $\Delta=\omega-\omega_0$ is the resonance detuning, $\gamma_{\perp}$ and $\gamma_{\parallel}$ are the transverse and longitudinal relaxation rates of the molecular transition, respectively, $\mu$ is the transition dipole moment, $D=\mathcal{N} w$, $w$ is the population inversion between the upper and lower molecular states and $D_0$ is the equilibrium value of $D$ determined by the optical pumping. The factor $1/3$ in front of $\mu^2$ in Eq. (\ref{eq:bloch1}) originates from the averaging over the orientations of molecules. \\
The field amplitudes ${\bf E}^{(\pm)}({\bf r},t)$ in the above equations should be regarded as the amplitudes of the local field in the layer of active molecules which depends itself on the molecular polarization. This dependence can be written in the form of an integral equation
\begin{equation}\label{eq:integral}
{\bf E}^{(+)}({\bf r},t)=\int^{\prime} \bar{\bf F}({\bf r},{\bf r}^{\prime}){\bf P}^{(+)}({\bf r}^{\prime},t)d{\bf r}^{\prime},
\end{equation}
where the prime above the integral sign implies removal of the point ${\bf r}^{\prime}={\bf r}$ from the integration. The kernel in this equation, $\bar{\bf F}({\bf r},{\bf r}^{\prime})$, is the so-called field susceptibility tensor which relates the electromagnetic field at the point ${\bf r}$ generated by a classical dipole with the dipole moment itself located at the point ${\bf r}^{\prime}$ \cite{Sipe84}.\\
In the bulk of the medium, Eq. (\ref{eq:integral}) is reduced to the Lorentz relation
\begin{equation}\label{eq:lorentz}
{\bf E}^{(+)}({\bf r},t)=\frac{4\pi}{3\epsilon_2}{\bf P}^{(+)}({\bf r},t).
\end{equation}
However at distances of the order of the wavelength or less from an interface the Lorentz field should be complemented by the dipole field reflected from the interface that can be written as
\begin{equation}
{\bf E}^{(+)}({\bf r},t)=\frac{4\pi}{3\epsilon_2}{\bf P}^{(+)}({\bf r},t)+\int \bar{\bf F}^R({\bf r},{\bf r}^{\prime}){\bf P}^{(+)}({\bf r}^{\prime},t)d{\bf r}^{\prime},
\end{equation}
where the superscript $R$ denotes the reflected field contribution.\\
\section{Lasing condition}\label{sec:lasing}
Let us assume now that initially at $t=0$ there is no electromagnetic field oscillating at the frequency $\omega$. We investigate the evolution of the system when there emerges a field of a small amplitude $\delta {\bf E}({\bf r},t)$. \\
In the linear approximation in $\delta {\bf E}({\bf r},t)$ one obtains the equations
\begin{eqnarray}\label{eq:linear}
\frac{\partial {\bf P}^{(+)}({\bf r},t)}{\partial t}=-(\gamma_{\perp }-i\Delta){\bf P}^{(+)}({\bf r},t)\nonumber\\
-\frac{i}{3\hbar } \mu^2D_0{\bf E}^{(+)}({\bf r},t)
\end{eqnarray}
and
\begin{eqnarray}
{\bf E}^{(+)}({\bf r},t)=\delta{\bf E}^{(+)}({\bf r},t)+\frac{4\pi}{3\epsilon_2}{\bf P}^{(+)}({\bf r},t)\nonumber\\
+\int \bar{\bf F}^R({\bf r},{\bf r}^{\prime}){\bf P}^{(+)}({\bf r}^{\prime},t)d{\bf r}^{\prime}.
\end{eqnarray}
Introducing the spatial Fourier transforms and the Laplace transforms in time,
\begin{equation}\label{eq:Efourier}
{\bf E}^{(+)}
({\bf k}_{\parallel},s;z)=\int\int_0^{\infty} {\bf E}^{(+)}({\bf r},t)e^{-i{\bf k}_{\parallel}\cdot {\bf r}_{\parallel}}e^{-st}dtd{\bf r}_{\parallel},
\end{equation}
\begin{equation}\label{eq:Pfourier}
{\bf P}^{(+)}({\bf k}_{\parallel},s;z)=\int\int_0^{\infty} {\bf P}^{(+)}({\bf r},t)e^{-i{\bf k}_{\parallel}\cdot {\bf r}_{\parallel}}e^{-st}dtd{\bf r}_{\parallel}
\end{equation}
with ${\bf r}_{\parallel}$ and ${\bf k}_{\parallel}$ being the radius vector and the wave vector components along the slab boundaries, respectively, one comes to the integral equation
\begin{eqnarray}\label{eq:integralFL}
\int_0^h\bar{\bf F}^R({\bf k}_{\parallel};z,z^{\prime}){\bf P}^{(+)}({\bf k}_{\parallel},s;z^{\prime})dz^{\prime}\nonumber\\
-\nu(s){\bf P}^{(+)}({\bf k}_{\parallel},s;z)=-\delta{\bf E}^{(+)}({\bf k}_{\parallel},s;z),
\end{eqnarray}
where 
\begin{equation}
\nu(s)=\frac{1-(4\pi/3\epsilon_2)\chi(s)}{\chi(s)}
\end{equation}
with
\begin{equation}
\chi(s)=-\frac{i}{3\hbar}\frac{\mu^2D_0}{s+\gamma_{\perp}-i\Delta},
\end{equation}
the quantity $\bar{\bf F}^R({\bf k}_{\parallel};z,z^{\prime})$ is determined by the Fourier transform
\begin{equation}\label{eq:Ffourier}
\bar{\bf F}^R({\bf r},{\bf r}^{\prime})=\frac{1}{(2\pi)^2}\int \bar{\bf F}^R({\bf k}_{\parallel};z,z^{\prime})e^{i{\bf k}_{\parallel}\cdot ({\bf r}_{\parallel}-{\bf r}^{\prime}_{\parallel})}d{\bf k}_{\parallel}
\end{equation}
and $\delta{\bf E}^{(+)}({\bf k}_{\parallel},s;z)$ is the spatial Fourier transform and the Laplace transform in time of the quantity  $\delta {\bf E}^{(+)}({\bf r},t)$.\\
The explicit form of the tensor $\bar{\bf F}^R({\bf k}_{\parallel};z,z^{\prime})$ is found in Ref. \cite{Sipe84} and can be written as
\begin{equation}\label{eq:FR}
\bar{\bf F}^R({\bf k}_{\parallel};z,z^{\prime})=\bar{\bf f}({\bf k}_{\parallel})e^{iq_2(k_{\parallel})(z+z^{\prime})}
\end{equation}
with 
\begin{equation}
q_2(k_{\parallel})=\sqrt{\frac{\omega^2}{c^2}\epsilon_2-k_{\parallel}^2},
\end{equation}
where the quantity $\bar{\bf f}({\bf k}_{\parallel})$ is given in the Appendix. This allows one to rewrite Eq. (\ref{eq:integralFL}) as follows
\begin{equation}\label{eq:matrix}
S(k_{\parallel})\hat{M}(k_{\parallel},s)\vec{P}(k_{\parallel},s)=-\delta\vec{E}(k_{\parallel},s),
\end{equation}
where
\begin{equation}
\vec{P}(k_{\parallel},s)=\int_0^h {\bf P}^{(+)}(k_{\parallel},s;z)e^{iq_2(k_{\parallel})z}dz,
\end{equation}
\begin{equation}
\delta\vec{E}(k_{\parallel},s)=\int_0^h \delta\mathbf{E}^{(+)}(k_{\parallel},s;z)e^{iq_2(k_{\parallel})z}dz,
\end{equation}
\begin{equation}
S(k_{\parallel})=\frac{1}{2iq_2(k_{\parallel})}\left[e^{2iq_2(k_{\parallel})h}-1\right]
\end{equation}
and the matrix $\hat{M}$ is defined as
\begin{equation}
\hat{M}(k_{\parallel},s)=\hat{f}(k_{\parallel})-\frac{\nu(s)}{S(k_{\parallel})}\hat{I}
\end{equation}
with $\hat{f}$ being the matrix of the tensor $\bar{\bf f}$ and $\hat{I}$ is the unit $3\times 3$ matrix. We have also taken into account that because of the axial symmetry of the problem nothing depends on the direction of the vector ${\bf k}_{\parallel}$.\\
As it follows from Eq. (\ref{eq:matrix}), the evolution of the polarization of active molecules is determined by the zeros of the determinant of the matrix $\hat{M}(k_{\parallel},s)$ which provide the poles of the polarization Laplace transform. On the other hand, these zeros are related with the eigenvalues  $\phi_j(k_{\parallel})$ of the matrix $\hat{f}(k_{\parallel})$ through the equation
\begin{equation}\label{eq:phi}
\phi_j(k_{\parallel})=\frac{\nu(s_j)}{S(k_{\parallel})},
\end{equation}
which implicitly determines the poles of the Laplace transform, $s_j(k_{\parallel})$. Resolving Eq. (\ref{eq:phi}) relatively $s_j$ one finds
\begin{equation}
s_j(k_{\parallel})=-\gamma_{\perp}+i\Delta^{\prime}+\frac{2\pi}{3\hbar}\mu^2D_0\mathcal{F}_j(k_{\parallel}),
\end{equation}
where
\begin{equation}
\Delta^{\prime}=\Delta-\frac{4\pi}{9\hbar\epsilon_2}\mu^2D_0,
\end{equation}
\begin{equation}
\mathcal{F}_j(k_{\parallel})=S(k_{\parallel})\tilde{\phi}_j(k_{\parallel})
\end{equation}
and the quantities $\tilde{\phi}_j(k_{\parallel})$ for $s$ and $p$ polarizations are given in the Appendix.\\ 
Now separating the real and imaginary parts of the poles, $s_j(k_{\parallel})=\sigma_j(k_{\parallel})+i\omega_j(k_{\parallel})$, one obtains
\begin{equation}\label{eq:sigma}
\sigma_j(k_{\parallel})=-\gamma_{\perp}+\frac{2\pi}{3\hbar}\mu^2D_0\text{Re}\mathcal{F}_j(k_{\parallel})
\end{equation}
and
\begin{equation}\label{eq:omega}
\omega_j(k_{\parallel})=\Delta^{\prime}+\frac{2\pi}{3\hbar}\mu^2D_0\text{Im}\mathcal{F}_j(k_{\parallel}).
\end{equation}
The positiveness of $\sigma_j(k_{\parallel})$ implies that the molecular polarization, and consequently the emitted field, grow exponentially with time that signifies the field generation (lasing). The lasing condition for a given polarization and propagation direction of the generated light can thus be found from Eq. (\ref{eq:sigma}) as follows
\begin{equation}\label{eq:threshold}
\frac{2\pi}{3\hbar}\mu^2D_0\text{Re}\mathcal{F}_j(k_{\parallel})>\gamma_{\perp}.
\end{equation}
On the other hand, Eq. (\ref{eq:omega}) determines the frequency of the generated field (the frequency pulling effect), 
\begin{equation}\label{eq:pulling}
\omega_g=\omega_0+\frac{4\pi}{9\hbar\epsilon_2}\mu^2D_0-\frac{2\pi}{3\hbar}\mu^2D_0\text{Im}\mathcal{F}_j(k_{\parallel}).
\end{equation}
Let us note that assuming a slow variation of the amplitude ${\bf P}^{(+)}(t)$ in time we implied a small detuning $\Delta$. Therefore the applicability criterion of Eq. (\ref{eq:pulling}) is $\mid \omega_g-\omega_0\mid\ll \omega_0$ that requires
\begin{equation}
\frac{2\pi}{3\hbar}\mu^2D_0\left\vert\text{Im}\mathcal{F}_j(k_{\parallel})\right\vert\ll\omega_0.
\end{equation}
The latter condition imposes also an upper limit for the left-hand side part of Eq. (\ref{eq:threshold}). This implies, in particular, that the developed theory is not applicable to the angular range close to the grazing propagation of the generated wave where the quantities $\tilde{\phi}_j(k_{\parallel})$ take very large values.\\
Let us introduce the dimensionless parameter
\begin{equation}
\eta=\frac{2\pi\mu^2D_0}{3\hbar\gamma_{\perp}}
\end{equation}
which characterizes the lasing threshold. We shall investigate the threshold condition, Eq. (\ref{eq:threshold}), in terms of the angle $\theta$ between the direction of the generated wave propagation and the normal to the slab boundaries which is related with $k_{\parallel}$ as follows
\begin{equation}
k_{\parallel}=\frac{\omega_g}{c}\sqrt{\epsilon_2}\sin\theta.
\end{equation}
Then the lasing condition takes the form
\begin{equation}
\eta>\frac{1}{\text{Re}\mathcal{F}_j(\theta)}.
\end{equation}
In the above consideration we assumed that there is a population inversion at the molecular transition that implies a positive $D_0$. However if the quantity $\text{Re}\mathcal{F}_j(\theta)$ is negative the lasing condition can be realized for a negative $D_0$. The criterion of such lasing without inversion is given by
\begin{equation}
\vert\eta\vert>\frac{1}{\left\vert\text{Re}\mathcal{F}_j(\theta)\right\vert}.
\end{equation}
In this case the pumping of the molecular transition is not necessary, but the number density of active molecules in the lower state should be large enough.  The phenomenon of lasing without inversion is based on a positive feedback provided by the slab with MNPs. This effect is phase-sensitive and its manifestation depends to a large extent on the average phase factor accumulated in the function $S(k_{\parallel})$.\\
It is worthwhile to note that, as it follows from Eq. (\ref{eq:threshold}), lasing is only possible if the quantities $\mathcal{F}_j$ are nonzero. In the case where $\epsilon_1\approx\epsilon_2\approx\epsilon_h$ and the dielectric slab does not contain MNPs the reflection coefficients $r^s_j$ and $r^p_j$ are close to zero (see the Appendix) and the same is true for the quantities $\tilde{\phi}_s$ and $\tilde{\phi}_p$. This makes the laser generation impossible that was observed in experiments \cite{Toropov21}.\\
\section{Numerical results}\label{sec:numerical}
\begin{figure}
\includegraphics[width=\linewidth]{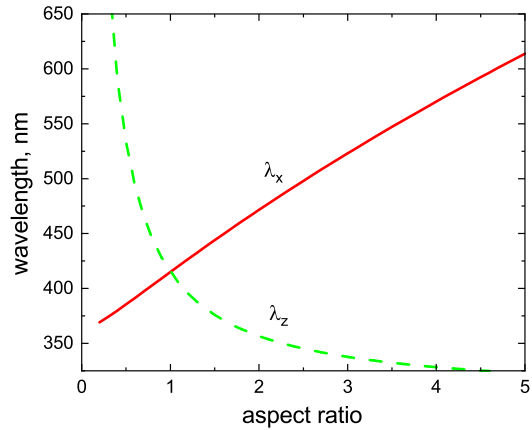}
\caption{\label{fig:frequency} The LSP wavelengths $\lambda_i=2\pi c/\omega_i$ of spheroidal Ag nanoparticles as functions of the aspect ratio  $\xi$.}
\end{figure}

\begin{figure}
\includegraphics[width=\linewidth]{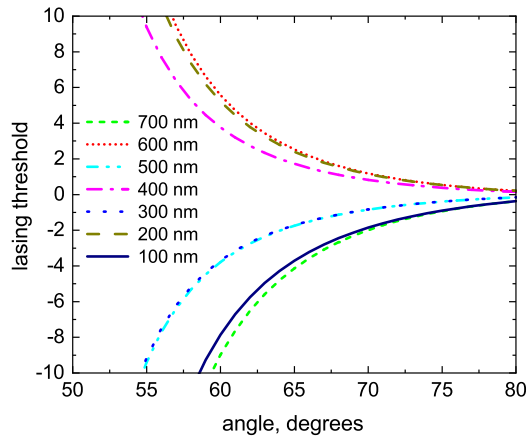}
\caption{\label{fig:threshold_p_h} The dimensionless lasing threshold in $p$ polarization, $1/\text{Re}\mathcal{F}_p(\theta)$, plotted for different values of the dye layer thickness, $h$, which are shown in the inset, $f=0.2$ and $d=70$ nm. The resonance condition $\omega_0\approx\omega_z$ is fulfilled for prolate spheroids with $\xi=0.5$. Only the range of the threshold parameter between $-10$ and $10$ is demonstrated.}
\end{figure}

\begin{figure}
\includegraphics[width=\linewidth]{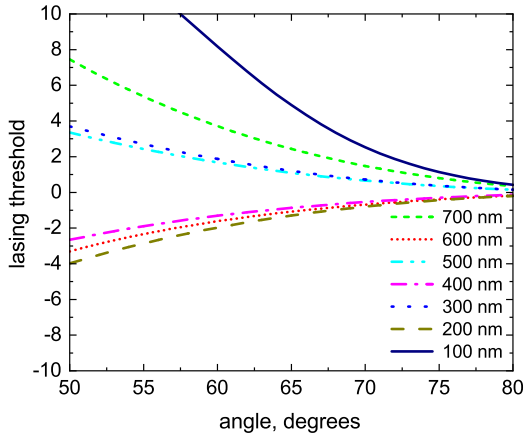}
\caption{\label{fig:threshold_s_h} Same as in Fig. \ref{fig:threshold_p_h}, but for $s$ polarization.}
\end{figure}

\begin{figure}
\includegraphics[width=\linewidth]{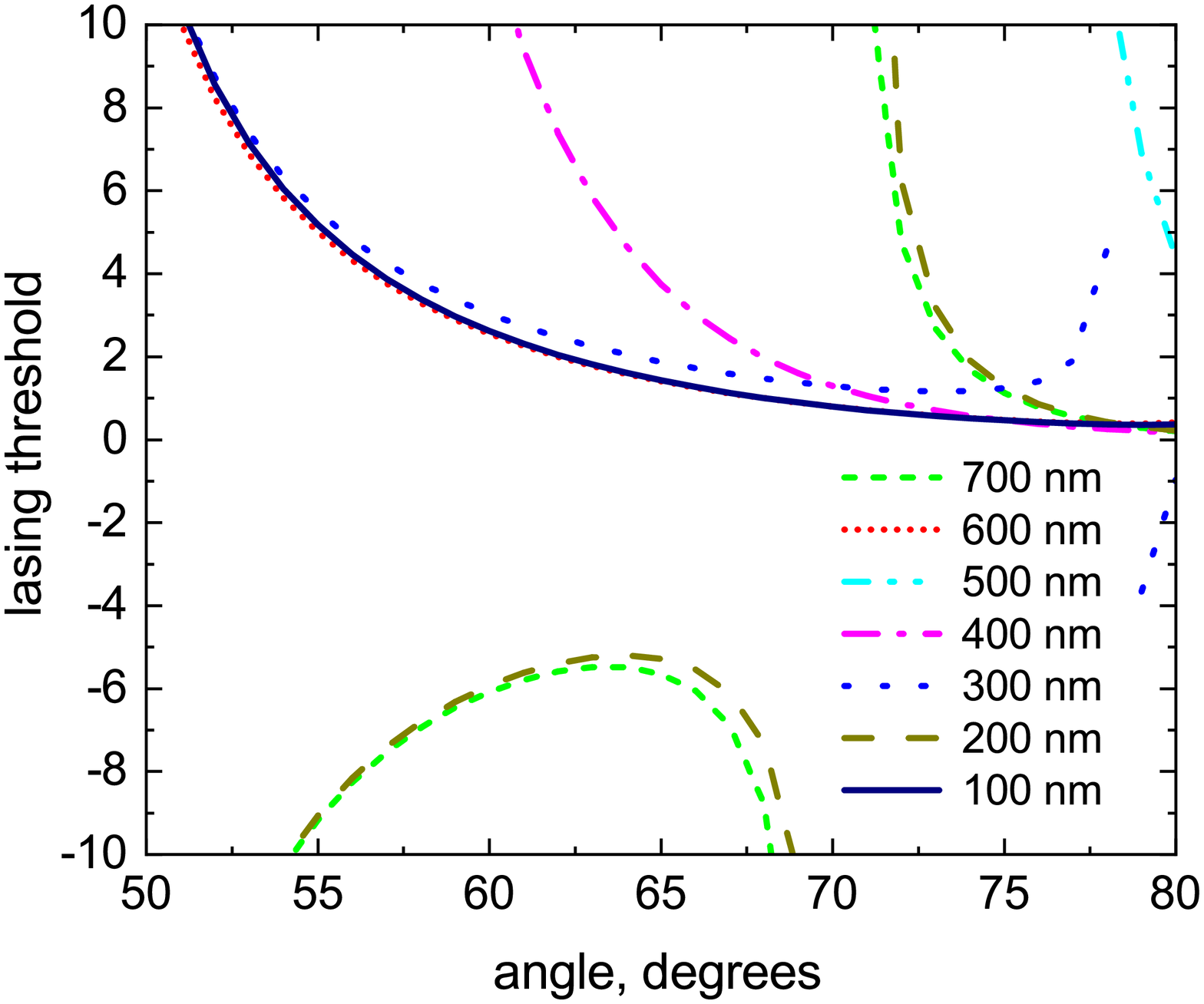}
\caption{\label{fig:threshold_p_h_x} Same as in Fig. \ref{fig:threshold_p_h}, but the resonance condition $\omega_0\approx\omega_x$ is fulfilled for oblate spheroids with $\xi=2.5$.}
\end{figure}

\begin{figure}
\includegraphics[width=\linewidth]{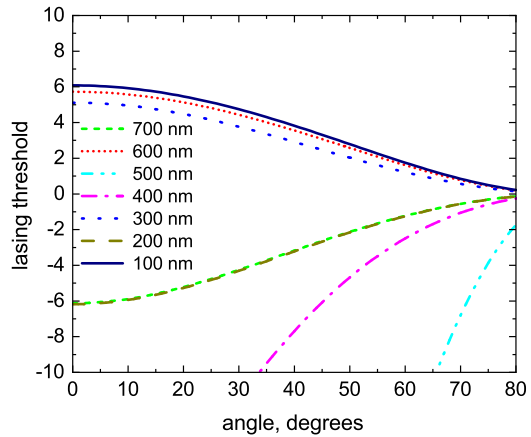}
\caption{\label{fig:threshold_s_h_x} Same as in Fig. \ref{fig:threshold_p_h_x}, but for $s$ polarization.}
\end{figure}

For the analysis of the lasing condition we assume that the dielectric functions of the dielectric materials, $\epsilon_1$, $\epsilon_2$ and $\epsilon_h$ are close to each other. On the one hand, this assumption corresponds to the experimental conditions realized in Ref. \cite{Toropov21}. On the other hand, it allows one to neglect the reflections from the dielectrics which are unimportant for the considered mechanism. We assume that the frequency of the active molecule transition is close to one of the LSP frequencies, i.e. either $\omega_0\approx\omega_x$ or $\omega_0\approx\omega_z$.\\
We illustrate the above developed theory by some numerical calculations performed for lasing in an array of Ag nanoparticles. The calculations have been carried out for the following values of parameters: $\epsilon_{\infty}=5$, $\omega_p=14.0\times 10^{15}$ s$^{-1}$, $\Gamma=0.32\times 10^{14}$ s$^{-1}$ \cite{Shalaev10} and $\epsilon_1=\epsilon_2=\epsilon_h=1.5^2$. Figure \ref{fig:frequency} shows the dependence of the LSP wavelengths $\lambda_x=\lambda_y$ and $\lambda_z$ on the aspect ratio of nanoparticles.\\
The typical values of the threshold parameter, $\eta$, achievable for laser dyes can be estimated as follows. Taking as an example the coumarin 481 dye used in Ref. \cite{Toropov21} and assuming its relative concentration in a solvent in the range $10^{-4}-10^{-2}$, one finds the number density of active molecules in the range  $\mathcal{N}\approx 3\times 10^{17}-3\times 10^{19}$ cm$^{-3}$. Assuming that the population inversion is maximal ($w=1$) and taking $\mu\approx 5$ D \cite{Nad03} and $\gamma_{\perp}\approx 10^{13}$ s$^{-1}$ \cite{Pustovit16}, one obtains $\eta\approx 0.002-0.2$. However if the solvent is evaporated during the dye layer deposition this quantity can take larger values.\\
While the lasing condition only weakly depends on the slab thickness $d$ and the MNPs volume fraction $f$ (not shown), the variation of the dye layer thickness $h$ leads to a dramatic change in the lasing threshold. Figure \ref{fig:threshold_p_h} demonstrates the dependence of the dimensionless threshold in $p$ polarization, $1/\text{Re}\mathcal{F}_p(\theta)$, for the case where $2\pi c/\omega_0\approx\lambda_z\approx 500$ nm that is fulfilled at $\xi=0.5$, i.e. for prolate spheroids. The threshold is invariant relatively the azimuthal angle and only the dependence on the polar angle $\theta$ is shown. One can see that for $h=200, 400$ and $600$ nm the threshold is positive, which means that conventional lasing can occur at polar angles where the value of $\eta$ is above the threshold. On the contrary, for $h=100, 300, 500$ and $700$ nm the threshold is negative and lasing without inversion is possible if the (negative) value of $\eta$ is below the threshold value. One can notice that the lasing condition does not change significantly if the value of $h$ is shifted by $200$ nm that allows one to reduce the size of the whole structure without loss in laser efficiency.\\
Figure \ref{fig:threshold_s_h} demonstrates a similar dependence, but calculated for $s$ polarization. Again one observes both positive and negative threshold values for different $h$, however, now the character of lasing (conventional versus lasing without inversion) is inverted for the same values of $h$. This implies also that lasing occurs in either $p$ or $s$ polarization that agrees with the observations reported in Ref. \cite{Toropov21}. In both cases lasing is possible in a limited angular range adjacent to the grazing propagation. \\
Figures \ref{fig:threshold_p_h_x} and \ref{fig:threshold_s_h_x} show the results of the calculations carried out for the case where the resonance condition $2\pi c/\omega_0\approx\lambda_x\approx 500$ nm is fulfilled for oblate spheroid with $\xi=2.5$. The conventional lasing in $p$ polarization can now be observed in the whole range of $h$, but within different angular ranges. Besides that, for $h=200$ and $700$ nm lasing without inversion is also possible. This time a $500$ nm shift in $h$ practically does not change the lasing condition. The same is true for $s$ polarization (Fig. \ref{fig:threshold_s_h_x}). However in the latter case the conventional lasing threshold becomes lower in the whole range of polar angles.
\section{Conclusion}\label{sec:conclusion}
In this paper, we have considered an array of spheroidal MNPs bordering with a layer of molecular gain medium. We have derived the lasing condition for such a structure as a function of the polar angle of the generated wave basing on a rigorous account of the local field in the molecular layer. \\
We have shown that the lasing threshold is determined by a product of the population inversion in molecules with the averaged phase factor which depends on the gain medium thickness. Such a condition predicts, besides conventional lasing, lasing without inversion if the phase factor is negative. For a given polar angle, the phase factor is a periodic function of the layer thickness that suggests a significant reduction of the overall structure size without loss in laser functionality.\\
The predicted possibility of lasing without inversion opens up a new avenue for the development of nanoscale coherent light sources which will not require powerful pumping and will be therefore low-cost energy-saving nanodevices. Another possible application of this effect is amplification of very weak nano-localized electromagnetic fields that can be exploited for ultra-sensitive surface chemical analysis. The findings of the present paper can stimulate further investigations in this promising direction.
\section*{Acknowledgments}
This research is kindly supported by the IE-Industrial Elektronik project (SFD-17-0036) which has received EU co-financing from the European Social Fund.

\appendix
\section{Field susceptibility tensor and its eigenvalues}
The tensor $\bar{{\bf f}}({\bf k}_{\parallel})$ introduced by Eq. (\ref{eq:FR}) has the form (see Ref. \cite{Sipe84})
\begin{equation}
\bar{{\bf f}}({\bf k}_{\parallel})=2\pi i\frac{\tilde{\omega}^2}{q_2(k_{\parallel})}\left[\hat{s}\hat{s}R^s(k_{\parallel})+\hat{p}_+\hat{p}_-R^p(k_{\parallel})\right],
\end{equation}
where $\tilde{\omega}=\omega/c$,
\begin{equation}
\hat{s}=\hat{k}_{\parallel}\times\hat{z}
\end{equation}
with $\hat{k}_{\parallel}$ and $\hat{z}$ being the unit vectors oriented along the corresponding directions,
\begin{equation}
\hat{p}_{\pm}=\left(\tilde{\omega}\sqrt{\epsilon_2}\right)^{-1}\left[k_{\parallel}\hat{z}\mp q_2(k_{\parallel})\hat{k}_{\parallel}\right],
\end{equation}
$R^s(k_{\parallel})$ and $R^p(k_{\parallel})$ are the Fresnel reflection coefficients for $s$ and $p$ polarizations, respectively.\\
The matrix of the tensor $\bar{{\bf f}}({\bf k}_{\parallel})$ in the basis of the vectors $\hat{s}$, $\hat{k}_{\parallel}$ and $\hat{z}$ takes the following form:
\begin{equation}
\hat{f}=2\pi i\left(\matrix{\frac{\tilde{\omega}^2}{q_2}R^s & 0 & 0 \cr
0 & -\frac{q_2}{\epsilon_2}R^p & -\frac{k_{\parallel}}{\epsilon_2}R^p \cr
0 & \frac{k_{\parallel}}{\epsilon_2}R^p & \frac{k_{\parallel}^2}{\epsilon_2q_2}R^p }\right).
\end{equation}
Correspondingly, its eigenvalues are found as $\phi_j(k_{\parallel})=2\pi i \tilde{\phi}_j(k_{\parallel})$ with
\begin{equation}\label{eq:phis}
\tilde{\phi}_s(k_{\parallel})=\frac{\tilde{\omega}^2}{q_2(k_{\parallel})}R^s(k_{\parallel}),
\end{equation}
\begin{equation}
\tilde{\phi}_{p1}(k_{\parallel})=0
\end{equation}
and 
\begin{equation}\label{eq:phip}
\tilde{\phi}_{p2}(k_{\parallel})=\frac{k_{\parallel}^2-q_2^2(k_{\parallel})}{\epsilon_2q_2(k_{\parallel})}R^p(k_{\parallel}).
\end{equation}
The reflection coefficients for a uniaxial slab disposed on a substrate are found as \cite{Lekner94}
\begin{equation}
R^s=\frac{r_2^s-r_1^s\exp{(2iq_od)}}{1-r_1^sr_2^s\exp{(2iq_od)}}
\end{equation}
and
\begin{equation}
R^p=\frac{r_2^p-r_1^p\exp{(2i\bar{q}d)}}{1-r_1^pr_2^p\exp{(2i\bar{q}d)}},
\end{equation}
where 
\begin{equation}
q_o=\sqrt{\left(\frac{\omega}{c}\right)^2\epsilon_o-k_{\parallel}^2},
\end{equation}
\begin{equation}
\bar{q}=\sqrt{\left(\frac{\omega}{c}\right)^2\epsilon_o-k_{\parallel}^2\frac{\epsilon_o}{\epsilon_e}},
\end{equation}
\begin{equation}
r_j^s=\frac{q_j-q_o}{q_j+q_o}
\end{equation}
and
\begin{equation}
r_j^p=\frac{Q-Q_j}{Q+Q_j}
\end{equation}
with $q_1=\sqrt{(\omega/c)^2\epsilon_1-k_{\parallel}^2}$, $Q_j=q_j/\epsilon_j$ and
\begin{equation}
Q=\frac{1}{\sqrt{\epsilon_o\epsilon_e} }\sqrt{\left(\frac{\omega}{c}\right)^2\epsilon_e-k_{\parallel}^2}.
\end{equation}


\begin{thebibliography}{99}
\bibitem{Odom17} A.~Yang, D.~Wang, W.~Wang, and T.W.~Odom, Annu. Rev. Phys. Chem. {\bf 68}, 83 (2017).
\bibitem{Zhang20} S.I.~Azzam, A.V.~Kildishev, R.-M.~Ma, C.-Z.~Ning, R.~Oulton, V.M.~Shalaev, M.I.~Stockman8, J.-L.~Xu, and X.~Zhang, Light Sci. Appl. {\bf 9}, 90 (2020).
\bibitem{Stockman03} D.J.~Bergman and M.I.~Stockman, Phys. Rev. Lett. {\bf 90}, 027402 (2003).
\bibitem{Hill07} M.T.~Hill, Y.-S.~Oei, B.~Smalbrugge, Y.~Zhu, T.~de~Vries, P.J.~van~Veldhoven, F.W.M.~van~Otten, T.J.~Eijkemans, J.P.~Turkiewicz, H.~de~Waardt, E.J.~Geluk, S.-H.~Kwon, Y.-H.~Lee, R.~N{\"o}tzel, and M.K.~Smit, Nat. Photonics {\bf 1}, 589 (2007). 
\bibitem{Noginov09} M.A.~Noginov, G.~Zhu, A.M.~Belgrave, R.~Bakker, V.M.~Shalaev, E.E.~Narimanov, S.~Stout, E.~Herz, T.~Suteewong, and U.~Wiesner, Nature {\bf 460}, 1110 (2009). 
\bibitem{Oulton09} R.F.~Oulton, V.J.~Sorger, T.~Zentgraf, R.-M.~Ma, C.~Gladden, L.~Dai, G.~Bartal, and X.~Zhang, Nature {\bf 461}, 629 (2009).
\bibitem{Berman94} P.R.~Berman (ed.), {\it Cavity Quantum Electrodynamics} (Academic Press, New York, 1994).
\bibitem{Walter06} H.~Walther, B.T.H.~Varcoe, B.-G.~Englert, and T.~Becker, Rep. Prog. Phys. {\bf 69}, 1325 (2006).
\bibitem{Drexhage68} K.H.~Drexhage, H.~Kuhn, and F.P.~Sch{\"a}fer, Ber. Bunsenges. Phys. Chem. {\bf 72}, 329 (1968).
\bibitem{Chance74} R.R.~Chance, A.~Prock, and R.~Silbey, J. Chem. Phys. {\bf 60}, 2744 (1974).
\bibitem{Bordo13} V.G.~Bordo, Phys. Rev. A {\bf 88}, 013803 (2013).
\bibitem{Pustovit16} V.N.~Pustovit, A.M.~Urbas, A.V.~Chipouline, and T.V.~Shahbazyan, Phys. Rev. B {\bf 93}, 165432 (2016).
\bibitem{Cuerda16} J.~Cuerda, F.J.~Garc{\'i}a-Vidal, and J.~Bravo-Abad, ACS Photonics {\bf 3}, 1952 (2016).
\bibitem{Shahbazyan17} T.V.~Shahbazyan, ACS Photonics {\bf 4}, 1003 (2017).
\bibitem{Bordo17a} V.G.~Bordo, Phys. Rev. B {\bf 95}, 235412 (2017).
\bibitem{Toropov21} N.~Toropov, A.~Kamalieva, A.~Starovoytov, S.~Zaki, and T.~Vartanyan, Adv. Photonics Res. {\bf 2}, 2000083 (2021).
\bibitem{Bordo16} V.G.~Bordo, Phys. Rev. B {\bf 93}, 155421 (2016).
\bibitem{Bordo17} V.G.~Bordo, Phys. Rev. A {\bf 96}, 023834 (2017).
\bibitem{Zheludev08} N.I.~Zheludev, S.L.~Prosvirnin, N.~Papasimakis, and V.A.~Fedotov, Nat. Photonics {\bf 2}, 351 (2008).
\bibitem{Landau} L.D.~Landau, E.M.~Lifshitz and L.P.~Pitaevskii {\it Electrodynamics of Continuous Media} (Elsevier, Amsterdam, 1984).
\bibitem{Haken} H.~Haken, {\it Light}, Vol. 2 (North-Holland, Amsterdam, 1985).
\bibitem{Sipe84} J.M.~Wylie and J.E.~Sipe, Phys. Rev. A {\bf 30}, 1185 (1984).
\bibitem{Shalaev10} W.~Cai and V.~Shalaev, {\it Optical Metamaterials} (Springer, New York, 2010).
\bibitem{Nad03} S.~Nad, M.~Kumbhakar, and H.~Pal, J. Phys. Chem. A {\bf 107}, 4808 (2003).
\bibitem{Lekner94} J.~Lekner, Pure Appl. Opt. {\bf 3}, 821 (1994). 
\end{thebibliography}
\end{document}